\documentstyle[secnumtab]{fbssuppl}

\input{tcilatex}

\begin{document}

\title{Significance Of Deuteron Breakup In A Halo Transfer Reaction}
\author{M. Yilmaz,and B. G\"{o}n\"{u}l 
\institute{Department of Engineering Physics,
University of Gaziantep,Gaziantep, 27310, Turkiye} \thanks{{\it Web address:}
http://www.gantep.edu.tr/~ayilmaz}}
\maketitle

\begin{abstract}
We discuss the quasi-adiabatic approximations to the three-body wavefunction
in breakup processes, clarifying the assumptions underlying the model. This
suggests alternative approximation schemes. Using different theoretical
three-body models, calculated differential cross section angular
distributions for the $^{11}$Be(p,d) reaction, for which new preliminary
data have been reported at 35 MeV, are presented. We show that calculations
are sensitive to the inclusion of deuteron breakup and to the breakup model
used, particularly if used to deduce absolute spectroscopic information on
the 0$^{+}$ and 2$^{+}$ $^{10}$Be core state parentages. There is also
considerable sensitivity to the model used in calculations of the relative
cross sections to the two states.
\end{abstract}

\section{Introduction}

In nuclei near the dripline, the separation energy of the last nucleon(s)
becomes extremely small. Compared with the common 6-8 MeV in stable nuclei,
many dripline nuclei have either a nucleon or a two-nucleon separation
energy that is less than 1 MeV. The neutron-density distribution in such
loosely bound nuclei shows an extremely long tail, called the neutron halo.
Although the density distribution of a halo is very low, it strongly affects
the reaction cross section and leads to new properties in such nuclei. The
study of halo nuclei is interesting as they involve new structures and
surface densities, and thus provide a stringent test of theoretical models
of nuclear structure and reactions.

Recently, the use of low energy single nucleon transfer reactions for
structure studies of exotic nuclei have attracted attention \cite{ref1}-\cite
{ref3}. Because of the simplicity of the theoretical interpretation of these
reactions, they are thought to provide an important source of the
information about the structure of halo nuclei.

Single nucleon transfer reactions, such as the (d,p) and (p,d) reactions,
have been a reliable tool in nuclear spectroscopic studies of stable nuclei,
determining positions, spins and parities of nuclear states. Developments in
radioactive beams technologies are now producing beams of nuclei at and near
the neutron and proton driplines, including neutron halo states with very
weak binding. The positions and the ordering of the nuclear single particle
levels in such systems, with large neutron excesses, has yet to be
clarified. These exotic systems typically have no bound excited states and
so traditional spectroscopic methods are inapplicable. Using the single
nucleon transfer process, on a deuteron or proton target, in an inverse
kinematics experiment, is then an attractive - although still very difficult
- alternative tool for mapping out nuclear structures near the driplines.

$^{11}$Be is a good example of such a system having a single neutron
separation energy of only 0.5 MeV. It is now understood that the 2s$_{1/2}$
neutron single particle state in this region is lowered and that a dominant
component of the $^{11}$Be ground state is produced by the coupling of a 2s$%
_{1/2}$ neutron to a $^{10}$Be (g.s., 0$^{+}$) core; with a smaller but
significant component in which a 1d$_{5/2}$ neutron is coupled to a $^{10}$%
Be (excited core, 2$^{+}$). This nucleus is of particular theoretical
interest because the ground state parity is exactly opposite to what one
naively would expect from the spherical shell model. There have been many
theoretical attempts to address the problem of parity inversion of the
ground state and the first excited state of the $^{11}$Be and evaluating the
extension of its neutron halo. Most of them \cite{ref4} correctly reproduce
the parity inversion but make very different predictions about the degree of
coupling of the $^{11}$Be ground state with the first 2$^{+}$ excited state
of $^{10}$Be at 3.368 MeV, the ratio of spectroscopic factors in these
different models $S$(2$^{+}$)/$S$(0$^{+}$) varying from 0.07 to 0.73. In
this context, the recent experiment at GANIL \cite{ref1} is expected to
measure the transfer reaction cross section angular distributions at forward
angles to the $^{10}$Be ground and 2$^{+}$ states, with the expectation of
clarifying quantitatively this $^{11}$Be (g.s.) admixture, and the
experimental determination of this ratio of spectroscopic factors by means
of a neutron pick-up reactions.

The purpose of the present work is to investigate the importance upon such
transfer reaction spectroscopic studies of the inclusion of the deuteron
breakup degrees of freedom in the theory used to analyze measured cross
section observables. Of importance in analyzing the experimental
measurements will be the extent to which the magnitudes of the calculated
cross sections, and particularly the ratio of the cross sections to the
ground state and 2$^{+}$ core states of $^{10}$Be, are affected by the
inclusion of three-body channels.

We discuss the $^{11}$Be (p,d) reaction within different theoretical
three-body (n+p+$^{10}$Be) models. We calculate the transfer amplitude using
the prior form of the (p,d) matrix element, thus the transition interaction
is the n-p interaction and we need a full (three-body) description of the
n+p+$^{10}$Be system in the final state. For the description of this final
state we have used

\begin{enumerate}
\item[i)]  the adiabatic (AD) model \cite{ref5},

\item[ii)]  the quasi-adiabatic (QAD) model \cite{ref6},\cite{ref7}, and

\item[iii)]  the projection operator (POA) approach \cite{ref8}.
\end{enumerate}

To clarify the importance of the breakup corrections to the transfer cross
sections, we also perform Distorted-Waves Born Approximation (DWBA)
calculations. For the ground state transition a preliminary test of the
models against more exact Coupled Discritized Continuum Channels (CDCC)
model \cite{ref9} calculations has been carried to provide an assessment of
the reliability of the three-body models used at the present incident energy.

In Section 2, we also present a re-formulation of the quasi-adiabatic model
for nuclear reactions which makes clear the approximations inherent in the
model. We show that the so called quasi-adiabatic breakup wavefunction also
contains the non-adiabatic corrections to the elastic part of the
wavefunction. This was a significant uncertainty in the formulations given
in Ref. \cite{ref6},\cite{ref7}. In addition, in the same section, an
alternative quasi-adiabatic scheme, the projection operator approach (POA),
based on the use of projection operators is introduced for a better
description of the non-adiabatic elastic channel contributions.

Current experimental activity in the area of light- neutron rich and
drip-line nuclei now dictates the rapid development of calculable
theoretical models for reactions and scattering of effective few-body
systems. Though, the CDCC model has been spectacularly successful in
understanding a wide range of data and phenomena in light- and heavy-ion
three-body systems, and can provide benchmark calculations against
approximate models such as AD, QAD and POA, it is unlikely ever to find an
application to the solution of many-body problems with more than three
interacting particles. Convergence problems may also become more serious
when very weakly bound halo nuclei are involved. To date such systems have
been analyzed using few-body Glauber based models \cite{ref10} whose first
step is an adiabatic treatment of the internal degrees of freedom of the
projectile. However, the application of the technique at lower projectile
energies, or to the case of including Coulomb breakup, has shown that the
model begins to break down due to known inadequacy of the adiabatic
approximation, see e.g. \cite{ref9},\cite{ref11}, at such energies. This
failure requires the treatment of non-adiabatic effects in the model.
Quasi-adiabatic ideas are therefore an obvious and necessary generalization
of these and non-eikonal adiabatic models. Hence, we first clarify the
formulation of quasi-adiabatic models within the three-body context in the
following Section. The calculation methods and the results of our
calculations are discussed in Section 4 and 5, respectively.

\section{The Three Body Models}

The n+p+nucleus Hamiltonian will be written

\begin{equation}
H=H_{np}+T_R+V_p(\overline{r}_p)+V_n(\overline{r}_n)+V_C(\overline{r}_p)
\label{eq1}
\end{equation}
where $H_{np}=(T_r+V_{np})$ is the n-p Hamiltonian, $T_R$ the center of mass
kinetic energy operator and $V_n$ and $V_p$ the neutron- and proton-target
effective interactions, regarded as local optical model potentials. $V_C$ is
the Coulomb field and is assumed to act on the n-p center of mass.

The total wavefunction $\Psi $ in the exit channel of the pick-up reaction
of interest satisfies

\begin{equation}
\left[ E-H_{np}-T_R-V(\overline{r},\overline{R})\right] \Psi (\overline{r},%
\overline{R})=0  \label{eq2}
\end{equation}

\begin{equation}
V(\overline{r},\overline{R})=V_p(\overline{r}_p)+V_n(\overline{r}_n)+V_C(%
\overline{R})  \label{eq3}
\end{equation}
where $\overline{r}$ is the relative coordinate of the n-p pair and $%
\overline{R}$ is the center of mass coordinate.

In the context of (p,d) reactions $\Psi $ enters the transition amplitude

\begin{equation}
T_{pd}=\left\langle \Psi (\overline{r},\overline{R})\right| V_{np}\left|
\chi ^{(-)}(\overline{r}_p)\phi _n(\overline{r}_n)\right\rangle  \label{eq4}
\end{equation}
with $\phi _n$ the neutron bound state and $\chi ^{(-)}$ incoming proton
wave function. Throughout this paper we restrict the formalism to S-wave n-p
relative motion for simplicity. In zero-range approximation then it is the
wavefunction at coincidence, $\Psi (0,\overline{R})$ which is of importance.

Theoretical methods based on the adiabatic treatment of one or more quantum
mechanical degrees of freedom have played an important role in nuclear
physics. More specifically, the Johnson-Soper application of adiabatic ideas 
\cite{ref5} to nuclear breakup effects played a key role in the development
of models of breakup processes in the three-body systems. However, in the AD
treatment it is assumed that the excitation of the projectile is to states
in the low energy continuum; its treatment of possible high energy breakup
contributions is therefore naturally suspect.

An improved treatment of these higher energy breakup configurations is
provided by the QAD method calculations which takes approximate account of
modifications to the center of mass energy of the n-p pair in breakup
configurations through the use of a mean breakup relative energy for the
continuum states. It thus breaks the degeneracy with the elastic channel.
The inclusion of these higher energy breakup components via the QAD model
calculations \cite{ref7} for large transferred angular momentum (d,p)
transfer reactions (at energies E/A $\approx $ 40 MeV) led to significant
deviations and systematic improvements over the adiabatic model in the
description of experimental data.

Although the quasi-adiabatic calculations produced an improved description
of the measured observables, the theoretical justifications of the
assumptions made in the model have not yet been studied. The work described
in this paper is thus concerned also with the clarification of these
theoretical uncertainties. In addition, we re-formulate the quasi-adiabatic
theory to give a more general formalism, approaching the three-body problem
in a different way. This alternative formulation provides a clear
understanding of the assumptions made in the original \cite{ref6}
quasi-adiabatic theory.

However, the quasi-adiabatic model does not include back-coupling
modifications to the elastic component of the wavefunction. For the
inclusion of these modifications, we also develop here an alternative
approximation scheme for the treatment of quantum mechanical three-body
systems using the unified theory of Feshbach.

\subsection{Adibatic Approach}

A considerable simplification in the solution of three-body equation is
achieved using the Johnson-Soper adiabatic approximation \cite{ref5}. The
approximation, in which the dynamics associated with the continuum coupling
collapses to an effective two-body problem for the n-p center of mass
motion, involves the replacement of $H_{np}$ in the exact three-body
Schr\"{o}dinger equation, Eq.(2), with $-\varepsilon _d$, the deuteron
binding energy. Therefore we have

\begin{equation}
\left[ E_{c.m.}-T_R-U(r,\overline{R})\right] \Psi ^{AD}(r,\overline{R})=0
\label{eq5}
\end{equation}
where $E_{c.m.}=E+\varepsilon _d$ is the energy of the outgoing deuteron in
the center of mass frame, and $U(r,\overline{R})$ is the angle average of $V(%
\overline{r},\overline{R})\ $ with respect to $\overline{r}$.

The assumption made is that the dominant breakup configurations are states
of low relative n-p energy $\varepsilon _k$ when compared $E$ with, such
that little error is made by assuming

\begin{equation}
E-\varepsilon k\cong E+\varepsilon _d  \label{eq6}
\end{equation}

In spite of the efficiency and success of the adiabatic approach,
experimental transfer reaction data \cite{ref12} clarified that some
physical contributions are missing from the calculation of the reaction
amplitude. These involve the transfer of particles in large n-p relative
energy configurations. As the adiabatic approximation is formulated under
the assumption of low energy n-p breakup, its treatment of high energetic
contributions is naturally suspect. An improved treatment of the higher
energy breakup configurations, such as quasi-adiabatic calculations, is thus
required.

\subsection{Quasi-adiabatic Model}

Under the restrictions to S-wave relative n-p configurations, one separates $%
\Psi $ into its elastic and breakup parts. Thus

\begin{equation}
\Psi ^{BU}(r,\overline{R})=\Psi (r,\overline{R})-\Psi ^{EL}(r,\overline{R})
\label{eq7}
\end{equation}
and therefore

\begin{equation}
\left[ E-H_{np}-T_R-U(r,\overline{R})\right] \Psi ^{BU}(r,\overline{R})=%
\left[ U(r,\overline{R})-U^{opt}(\overline{R})\right] \Psi ^{EL}(r,\overline{%
R})  \label{eq8}
\end{equation}
where $U^{opt}\Psi ^{EL}=$ $\left[ E_{c.m.}-T_R\right] \Psi ^{EL}$. At this
stage, the quasi-adiabatic model assumes:

\begin{enumerate}
\item  that $H_{np}$ is replaced by an average energy depending at most
parametrically upon $r$, and
\end{enumerate}

that the elastic part of the wavefunction generated by the adiabatic model $%
\Psi ^{AD,EL}$ is an accurate representation of $\Psi ^{EL}$. Thus the
quasi-adiabatic approximation to $\Psi ^{BU}$ solves 
\begin{equation}
\left[ E-\overline{\varepsilon }-T_R-U(r,\overline{R})\right] \Psi
^{QAD,BU}(r,\overline{R})=\left[ U(r,\overline{R})-U^{AD,opt}(\overline{R})%
\right] \Psi ^{AD,EL}(r,\overline{R})  \label{eq9}
\end{equation}
where $\overline{\varepsilon }$ is taken \cite{ref7} as the expectation
value of $H_{np}$ in $\Psi ^{AD,BU}$ in each partial wave, and $%
U^{AD,opt}\Psi ^{AD,EL}=\left[ E_{c.m.}-T_R\right] \Psi ^{AD,EL}$.

Thus, the quasi-adiabatic approximation removes the degeneracy of the n-p
center-of-mass energy in breakup configurations by introducing a positive
mean energy for the continuum states. We note that the model provides,
however, no prescription for modifications to the elastic component of the
wavefunction.

\subsection{An Alternative Formulation Of The Quasi-adiabatic Model}

We present here an alternative development of a formal theory for the
quasi-adiabatic method, which clarifies that such a model can be introduced
by making only one single approximation, unlike the formulations of the
model in Ref. \cite{ref6},\cite{ref7}. Another advantage of this formulation
is that one sees how to treat corrections to both the elastic and breakup
components of the wavefunction, and to derive an iterative scheme for such
changes. This was a significant uncertainty in the original quasi-adiabatic
formulation of Amakawa et al. \cite{ref6}, in which it is stated that the
elastic wavefunction is assumed unchanged, regardless of changes made in the
breakup piece of the wavefunction.

Under the restriction to S-wave relative n-p configurations, a formal
development of the quasi-adiabatic theory proceeds by decomposing the
projectile-target three-body wavefunction into the adiabatic wavefunction
plus the correction term, i.e.,

\begin{equation}
\Psi (r,\overline{R})=\Psi ^{AD}(r,\overline{R})+\Delta \Psi (r,\overline{R})
\label{eq10}
\end{equation}
where $\Delta \Psi $ accounts for non-adiabatic corrections to the breakup
and elastic channels, and has only outgoing waves since $\ \Psi ^{AD}$
already satisfies incoming wave boundary conditions. Upon substitution in
the Schr\"{o}dinger equation then,

\begin{equation}
\left[ E-H_{np}-T_R-U(r,\overline{R})\right] \Delta \Psi (r,\overline{R}%
)=(H_{np}+\varepsilon _d)\Psi ^{AD,BU}(r,\overline{R})  \label{eq11}
\end{equation}
where the source term has infinite range. To proceed we use the
inhomogeneous equation for $\Psi ^{AD,BU}$,

\begin{equation}
\Psi ^{AD,BU}(r,\overline{R})=\left[ E_{c.m.}-T_R-U(r,\overline{R})\right]
^{-1}\left[ U(r,\overline{R})-U^{AD,opt}(\overline{R})\right] \Psi
^{AD,EL}(r,\overline{R})  \label{eq12}
\end{equation}
from which

\begin{eqnarray}
\Delta \Psi (r,\overline{R}) &=&\left[ E-H_{np}-T_R-U(r,\overline{R})\right]
^{-1}(H_{np}+\varepsilon _d)\left[ E_{c.m.}-T_R-U(r,\overline{R})\right]
^{-1}  \label{eq13} \\
&&\times \left[ U(r,\overline{R})-U^{AD,opt}(\overline{R})\right] \Psi
^{AD,EL}(r,\overline{R})  \nonumber
\end{eqnarray}

It follows that

\begin{eqnarray}
\Delta \Psi (r,\overline{R}) &=&\left\{ \left[ E-H_{np}-T_R-U(r,\overline{R})%
\right] ^{-1}-\left[ E_{c.m.}-T_R-U(r,\overline{R})\right] ^{-1}\right\}
\label{eq14} \\
&&\times \left[ U(r,\overline{R})-U^{AD,opt}(\overline{R})\right] \Psi
^{AD,EL}(r,\overline{R})  \nonumber
\end{eqnarray}
and hence we can write $\Delta \Psi (r,\overline{R})=\Delta \Psi _1(r,%
\overline{R})-\Delta \Psi _2(r,\overline{R})$ with

\begin{equation}
\left[ E-H_{np}-T_R-U(r,\overline{R})\right] \Delta \Psi _1(r,\overline{R})=%
\left[ U(r,\overline{R})-U^{AD,opt}(\overline{R})\right] \Psi ^{AD,EL}(r,%
\overline{R})  \label{eq15}
\end{equation}

\begin{equation}
\left[ E_{c.m.}-T_R-U(r,\overline{R})\right] \Delta \Psi _2(r,\overline{R})=%
\left[ U(r,\overline{R})-U^{AD,opt}(\overline{R})\right] \Psi ^{AD,EL}(r,%
\overline{R})  \label{eq16}
\end{equation}

The equation for $\Delta \Psi _1$ leads to the original quasi-adiabatic
wavefunction. $\Delta \Psi _2$, precisely the adiabatic breakup wavefunction
and has no overlap with the elastic channel. Thus $\Delta \Psi _1$ must
include both elastic and breakup non-adiabatic corrections. The elastic
piece can be extracted by projection. This makes clear that the assumption $%
\Psi ^{EL}$ $\approx \Psi ^{AD,EL}$ of the original formulation is
unnecessary. Therefore the reduction of the exact three-body equation to the
quasi-adiabatic model requires only the replacement of $H_{np}$ by an
average breakup energy $\overline{\varepsilon }$. The details of the similar
resulting partial wave expansions and solution of the equations can be found
in Sec. III of Ref. \cite{ref7}.

\subsection{Projection-Operator Approach}

In the previous analyses, the non-adiabatic elastic corrections are clearly
treated incorrectly. They solve equations with the wrong asymptotic energy.
In case of deuteron scattering the non-adiabatic elastic corrections are
small but in general this may not be the case. A more correct treatment
requires use of projection operators on $P$ and off $Q$ the projectile
ground state wavefunction,

\begin{equation}
\Psi =P\Psi +Q\Psi =\Psi ^{EL}+\Psi ^{BU}  \label{eq17}
\end{equation}

The usual reduction of the Schr\"{o}dinger equation can be shown \cite{ref8}
to yield coupled equations

\begin{equation}
\left[ E-QH_{np}Q-T_R-U(r,\overline{R})\right] \Psi ^{BU}(r,\overline{R})=%
\left[ U(r,\overline{R})-U^{opt}(\overline{R})\right] \Psi ^{EL}(r,\overline{%
R})  \label{eq18}
\end{equation}
\begin{equation}
\left[ E_{c.m.}-T_R-U_{dd}(\overline{R})\right] \Psi ^{EL}(r,\overline{R}%
)=\left| \phi _d(r)\right\rangle \left\langle \phi _d(r)\right| U(r,%
\overline{R})\left| \Psi ^{BU}(r,\overline{R})\right\rangle  \label{eq19}
\end{equation}
with $U_{dd}$ the Watanabe potential and $\phi _d$ the deuteron ground state
wavefunction. Again at this stage $H_{np}$ is to be replaced with an average
excitation energy $\overline{\varepsilon }$, in each partial wave. The
equations can be solved by iteration \cite{ref8} starting from the adiabatic
model estimates.

\section{Experimental Aspect Of The Reaction}

In the past, transfer reactions with light and heavy ions have been a major
source of spectroscopic information in stable nuclei, including spin and
parity assignments to nuclear levels, measurements of occupation
probabilities and wave functions in the ground and excited states of the
daughter system. Provided that beams of exotic nuclei with high enough
luminosity will become available it is tempting to use transfer reactions in
the same spirit also for spectroscopic studies in exotic nuclei. The
important difference to former fixed target experiments is the low intensity
of secondary beams and the use of inverse kinematics. Under these conditions
only transfer reactions with rather large cross sections can be investigated.

Since exotic beams are obtained as secondary beams they depend on the
production rates for unstable nuclei in primary reactions with a stable
beam. As a consequence, their intensities are reduced by several orders of
magnitude with respect to the intensities of the primary beam. As a rough
rule, the production cross sections for these beams fall by approximately
one order of magnitude for each mass unit going further away from stability 
\cite{ref2}. Thus, one is restricted to reactions with larger cross
sections. Targets should be chosen as thick as possible and detectors with
large solid angles and high detection efficiency should be used.

Obviously, the abundance of data to which one is accustomed from work with
stable beams can certainly not be expected for radioactive beams. In
reactions with inverse kinematics the cross-sections are focussed more in
forward direction in the laboratory system. Thus, the full yield of the
nucleus to be investigated is collected in a rather small angular cone and
total cross section measurements should already be feasible at much lower
beam intensities.

As a representative example, we consider here an inverse kinematics reaction
with beam of neutron rich nucleus $^{11}$Be. The $^{11}$Be (p,d) $^{10}$Be
reaction, leading to bound states in $^{10}$Be, has been studied for the
first time \cite{ref1},\cite{ref2}, using a secondary $^{11}$Be beam of 35
MeV/nucleon. Angular distributions up to about 15 degrees (in c.m.) were
measured by detecting $^{10}$Be in a spectrometer and coincident deuterons
in a position sensitive silicon detector array. Their preliminary analysis
provides evidence for a large core excitation component in the structure of $%
^{10}$Be ground state.

\section{Calculations}

In the present work, the $^{11}$Be system is interpreted as a $^{10}$Be
core, with one excitable state, and a loosely bound neutron. We have only
included the 0$^{+}$ ground state and 2$^{+}$ excited state of the core.
That means, in the ground state of $^{11}$Be, the neutron oscillates between
a 2s$_{1/2}$ and 1d$_{5/2}$ orbital, exciting the core from a 0$^{+}$ to a 2$%
^{+}$ state (and de-exciting it).

Since a proper model provides modifications to the three-body
deuteron-channel wavefunction, one expects the calculations resulting from
this method to improve upon the predictions on reaction observables. Thus,
in order to establish the significance of the modifications introduced by
the models, and for future comparison to the experimental data, we have
evaluated differential cross-section angular distributions at 35 MeV
incident energy for the $^{11}$Be (p,d) $^{10}$Be (g.s and 2$^{+}$)
transitions and the trends of these observables have been compared with each
other.

Three-body wavefunctions in the exit channel of the reaction are calculated
using the adiabatic, quasi-adiabatic, projection operator and CDCC methods
(the later using the code FRESCO \cite{ref13}). The main objective of the
CDCC calculations, which do not include spin-orbit interactions due to the
difficulty in including this term in FRESCO, is to provide a critique of the
theories implemented here, rather than to seek a realistic comparison with
experiment. The preliminary test (for 2s$_{1/2}$ transition) of the AD and
extended AD models using the CDCC calculations, discretizing the n-p
continuum into 6 bins from 0 to 25 MeV excitation energy, provide an insight
into their reliability at the lower incident proton energy of interest.

The quasi-adiabatic and projection operator model calculations are iterated
in the sense that the mean breakup energy for the continuum, $\overline{%
\varepsilon }_{JL}$, in each partial wave is calculated from the latest best
estimate of the breakup wavefunction, i.e.

\begin{equation}
\overline{\varepsilon }_{JL}^{(i)}(R)=\frac{\left\langle ^{(i)}\Psi
_{JL}^{BU}(r,\overline{R})\right| H_{np}\left| ^{(i)}\Psi _{JL}^{BU}(r,%
\overline{R})\right\rangle _r}{\left\langle ^{(i)}\Psi _{JL}^{BU}(r,%
\overline{R})\mid ^{(i)}\Psi _{JL}^{BU}(r,\overline{R})\right\rangle _r}
\label{eq20}
\end{equation}
where bra-ket denotes the radial integration over $r$ and $(i)$ represents
the iteration number. Although this prescription reproduces meaningful
breakup energies at medium incident projectile energies for heavy targets 
\cite{ref8}, it breaks down at lower incident energies and cannot be
iterated due to the calculated unphysical $\overline{\varepsilon }_{JL}$
values which are larger than the center of mass energy. To overcome this
problem for the underlying reaction, we consider another formulation for the
mean breakup energies.

Starting with an exact definition of the continuum channel breakup
wavefunction in partial wave form ,

\begin{equation}
\Psi _{JL}^{BU}(r,\overline{R})=\int_0^\infty dk\phi _k(r)\chi _{JLk}(R)
\label{eq21}
\end{equation}
where $\phi _k(r)$ is a triplet n-p scattering state with asymptotic
normalization

\begin{equation}
\phi _k(r)\longrightarrow \left( \sqrt{\frac 2\pi }\right) \frac{\sin
(kr+\delta _0)}r  \label{eq22}
\end{equation}
such that $\int_0^\infty drr^2\phi _k(r)\phi _{k^{\prime }}(r)=\delta
(k-k^{\prime })$, one can rearrange the mean energy prescription, Eq. (\ref
{eq20}), in the form

\begin{eqnarray}
\overline{\varepsilon }_{JL}(R) &=&\frac{\left\langle \Psi _{JL}^{BU}\left|
H_{np}\right| \Psi _{JL}^{BU}\right\rangle _r}{\left\langle \Psi
_{JL}^{BU}\right| \Psi _{JL}^{BU}\rangle _r} \\
&=&\frac{\left\langle \int_0^\infty dk\phi _k(r)\chi _{JLk}(R)\right|
\int_0^\infty dk\varepsilon _k\phi _k(r)\chi _{JLk}(R)\rangle _r}{%
\left\langle \int_0^\infty dk\phi _k(r)\chi _{JLk}(R)\right| \int_0^\infty
dk\phi _k(r)\chi _{JLk}(R)\rangle _r} \\
&=&\frac{\int_0^\infty dk\chi _{JLk}^{*}(R)\varepsilon _k\chi _{JLk}(R)}{%
\int_0^\infty dk\chi _{JLk}^{*}(R)\chi _{JLk}(R)}  \nonumber
\end{eqnarray}
where * denotes complex conjugate and $H_{np}\phi _k(r)=\varepsilon _k\phi
_k(r)$ in which $\varepsilon _k=\frac{\hbar ^2k^2}{2\mu _{np}}$. One needs
at this stage to evaluate the $\chi _{JLk}$, by integration

\begin{equation}
\chi _{JLk}(R)=\int_0^\infty drr^2\phi _k(r)\Psi _{JL}^{BU}(R)  \label{eq24}
\end{equation}
where $\Psi _{JL}^{BU}$ is calculated approximately by the QAD and POA
theories. To be in consistent with the CDCC calculations carried out in this
work, we set the maximum value of $k$ for the integral in Eq. (23) to 0.78 fm%
$^{-1}$ that corresponds to 25 MeV for the relative breakup energies.

The calculated $\overline{\varepsilon }_{JL}$ values show significant
changes from the zeroth order estimate based on $\Psi ^{AD,BU}$ but beyond
the next iteration such changes are not reflected in changes in the
calculated wavefunctions or in reaction observables.

The calculated three-body wavefunctions using the same inputs, in partial
waveform, provided by the present models are employed in a modified version
of the program TWOFNR \cite{ref14} for the evaluation for the reaction
observables, performing the zero-range approximation. The zero-range
approximation is expected to produce a more accurate model for in particular
2s$_{1/2}$ state since there is a significant probability of the neutron in $%
^{11}$Be being near the $^{10}$Be core. Also, comparisons with full finite
range calculations in the DWBA and CDCC cases, where the n-p interaction was
taken to be a central Hulthen potential, showed this to be a reasonable
first approximation.

The entrance and exit channel potentials are obtained from the global
parameterization of Bechetti and Greenless \cite{ref15}. We note however at
this point that for the rigorous reproduction of experimental data, one
needs to consider different combinations of optical potentials for the
proton and deuteron channels in the calculations in order to test the
sensitivity of extracted spectroscopic factors to the input parameters.
Despite the uncertainties on optical potentials used in our calculations,
which in fact are important for a precise description of transfer cross
sections, the results can however be expected to describe realistically the
dynamical features of such reactions. A clear improvement for future work
and the analysis of data would be achieved by measuring elastic scattering
together with transfer reactions such that empirical information on optical
potentials is obtained.

The radial integrals are carried out from 0 to 35 fm in steps of 0.1 fm. The
maximum number of partial waves used is 30 for both entrance and exit
channels. The transferred neutron wavefunction is evaluated in a Woods-Saxon
well with shape parameters $r$=1.25 fm and $a$=0.65 fm. The real well depth
is adjusted to reproduce the neutron separation energy. The spin-orbit force
in the proton channel is fixed at 6 MeV. The spectroscopic factors are set
to unity throughout the calculations.

All calculations presented here are done without non-locality corrections.
Such corrections for halo transfer are expected to be small \cite{ref3}
because they correct the transition amplitude in the nuclear interior, but
the long tail of the halo wavefunction makes internal contributions less
important.

\section{Results And Discussion}

From the results shown in Figs.\ref{Fig 1.} and \ref{Fig 1.}, it is clear
that the inclusion of deuteron breakup is of importance for extracting
information on the 0$^{+}$ and 2$^{+}$ states. The calculated cross-section
angular distributions also show significant differences when using the
different breakup treatments. In addition, cross-sections of the $^{11}$%
Be(p,d)$^{10}$Be (g.s., 0$^{+}$) reaction calculated with different
three-body models are larger but decrease faster at forward angles when
compared with the calculations using DWBA which do not account for the
effects arising from the breakup of the deuteron in the field of the
nucleus. To provide an assessment of the three-body models (including
spin-orbit interactions) used here we have compared in Fig. \ref{Fig 1.} the
calculations (for the 2s$_{1/2}$ transition) with those obtained using the
CDCC technique (excluding spin-orbit forces). Based on this comparison, the
AD, QAD and POA are reliable tools for the spectroscopic study of the
reaction of interest. The angular distributions of the transfer reactions to
the 2$^{+}$ state of $^{10}$Be at small angles look the same both with and
without breakup effects but have different absolute values.

Our calculations have demonstrated that the inclusion of breakup effects in
transfer reactions produces changes in the shape of the angular
distributions and these effects increase the absolute values of the
theoretical cross-sections in forward direction and thus lead to the smaller
values of the spectroscopic factors extracted from the experimental data. It
is expected that such effects will be larger when the incident energy
increases and the mass of the target decreases \cite{ref2},\cite{ref3}.

We therefore conclude that the standard procedure for the determination of
spectroscopic factors as ratios of the experimental transfer cross-sections
to those calculated within the standard DWBA is not reliable for the
reactions involving weakly bound halo nuclei. The spectroscopic factors
extracted from the experimental data would be smaller than those obtained
with the conventional DWBA, due to including breakup effects.

Our results also suggest that deuteron breakup stronger for 2s$_{1/2}$
transfer than for 1d$_{5/2}$ transfer. This may be associated with the node
in the bound state wavefunction in the 2s$_{1/2}$ case. In the case of the 2s%
$_{1/2}$ transfer the shape of the differential cross-sections changes more
strongly at small angles compared to the 1d$_{5/2}$ transfer. This should
influence the ratio of the spectroscopic factors for 0$^{+}$and 2$^{+}$
states of the$^{10}$Be obtained with different theoretical models.

Fig. \ref{Fig 3.} shows the ratio of the calculated cross-sections relevant
to deducing only relative spectroscopic information. The figure leads us to
the conclusion that ratios of spectroscopic factors depend on possible
uncertainties on absolute cross-section values, and are typically dependent
upon the ingredients of reaction calculations with respect to scattering
angles.

In the broad field of today's nuclear structure research , there is still
very much to be understood when we analyze light exotic nuclei. Our
intention for the near future is to have a better insight into the physical
nature of the less known halo systems, using the models employed in this
work. In particular, the question about the halo nature of $^{19}$C and its
underlying structure is one of the interesting current questions in dripline
physics. More information on this carbon isotope will allow us to further
explore the characteristics of the halo phenomenon and to test the concepts
developed for $^{11}$Be.

\newpage\ 

\begin{center}
{\bf Acknowledgments}
\end{center}

The financial support of the Scientific and Technical Research Council in
T\"{u}rkiye (T\"{U}BITAK) in the form of Grant No. TBAG 1601 is gratefully
acknowledged. One of the authors (M.Y.) also thanks to the British Council
in Turkiye and to the Physics Department at University of Surrey in UK for
providing a short visit to Surrey University to carry out and discuss some
part of the calculations. Enlightening discussions with Dr J. A. Tostevin at
Surrey University are gratefully acknowledged.

\newpage\

\newpage\ 

\begin{center}
{\bf Figure Captions}
\end{center}

Fig \ref{Fig 1.} Calculated differential cross-section angular distributions
for the $^{11}$Be(p,d)$^{10}$Be (g.s.) reaction at 35 MeV using different
theoretical models. The spectroscopic factor is 1.0 for all calculations.

Fig. \ref{Fig 2.} As for Fig. \ref{Fig 1.}, but for the $^{11}$Be(p,d)$^{10}$%
Be (2$^{+}$, 3.368 MeV) state transition.

Fig. \ref{Fig 3.} The ratio of the calculated relative cross-sections to the
2$^{+}$ and 0$^{+}$ $^{10}$Be final states.


\begin{thebibliography}{10}
\bibitem[1]{ref1}  J.S. Winfield et al., presented at: Nuclear Structure at
the Extremes, Lewes, Sussex, U.K., June 17-19 1998, to be published in
J.Phys.G: Nucl.Part.Phys.; S. Fortier et al., presented at: 2nd
International Conference on Exotic Nuclei and Atomic Masses - ENAM98, Shanty
Creek, Michigan, June 23-27 1998, AIP Conference series, in the press.

\bibitem[2]{ref2}  H. Lenske and G. Schrieder, Eur. Phys. J. A, {\bf 2}
(1998) 41.

\bibitem[3]{ref3}  N. K. Timofeyuk and R. C. Johnson, Surrey Univ. Report
SCNP-98/12, submitted to Phys. Rev. {\bf C}.

\bibitem[4]{ref4}  E. K. Warburton and B. A. Brown, Phys. Rev. {\bf C46}
(1992) 923; T. Otsuka et al Phys. Rev. Lett {\bf 70} (1993) 1385; P.
Descouvemont, Nucl. Phys. {\bf A615} (1997) 261; N. Vinh Mau, Nucl. Phys. 
{\bf A592} (1995) 33; T. Bhattacharya and K. Krishan, Phys. Rev. {\bf C56 }%
(1997) 212; H. Esbensen, B. A. Brown and H. Sagawa, Phys. Rev. {\bf C31}
(1995) 1274; F. M. Nunes, I. J. Thompson and R. C. Johnson, Nucl. Phys. {\bf %
A592} (1995) 33.

\bibitem[5]{ref5}  R. C. Johnson and P. J. R Soper, Phys. Rev. {\bf C1}
(1970) 976.

\bibitem[6]{ref6}  H. Amakawa, N. Austern, and C. M. Vincent, Phys. Rev. 
{\bf C29} (1984) 699.

\bibitem[7]{ref7}  E. J. Stephenson et al., Phy. Rev. {\bf C42} (1990) 2562.

\bibitem[8]{ref8}  J. A. Tostevin, and B. Gonul, Genshikaku Kenkyu, {\bf %
Vol. 40} No.2, (1995) 95; B. G\"{o}n\"{u}l, PhD Thesis, Department of
Physics, University of Surrey, Guildford, England, 1994.

\bibitem[9]{ref9}  M. Kamimura et al., Prog. Theor. Phys. Suppl. {\bf 89}
(1986) 1; N. Austern et al., Phys. Rep. {\bf 154} (1987) 125.

\bibitem[10]{ref10}  K. Yabana et al., Phy. Rev. {\bf C45} (1992) 2909; I.
J. Thompson et al., Phy. Rev. {\bf C47}, R1364 (1993); J. S. Al-Khalili, I.
J. Thompson and J. A. Tostevin, Nucl. Phys. {\bf A581} (1995) 331; J. A.
Tostevin and J. S. Al-Khalili, Phys. Rev. {\bf C56} R2929 (1997); J. S.
Al-Khalili and J. A. Tostevin, Phys. Rev. {\bf C57} (1998) 1846.

\bibitem[11]{ref11}  H. Amakawa and N. Austern, Phys. Rev. {\bf C27} (1983)
922; G. H. Rawitscher, Phys. Rev. {\bf C11} (1975) 1152.

\bibitem[12]{ref12}  E. J. Stephenson et al., Phys. Lett. {\bf B171} (1986)
358; E. J. Stephenson et al., Nucl. Phys. {\bf A469} (1987) 467; R. C.
Johnson, E. J. Stephenson, and J. A. Tostevin, {\bf A505} (1989) 26.

\bibitem[13]{ref13}  I.J.Thompson, Computer program FRESCO,University of
Surrey, unpublished;Computer Physics Report, {\bf 7} (1988) 167.

\bibitem[14]{ref14}  M. Igarashi, M. Toyoma and N. Kishida, Computer Program
TWOFNR (Surrey University version ).

\bibitem[15]{ref15}  F. D. Bechetti and G. W. Greenlees, Phys. Rev. {\bf 182}%
, 1190 (1969).
\end{thebibliography}
\end{document}